# On the frequency of the superfireballs: more than 150 years of reports

Sandra Zamora[1], Francisco Ocaña[1], Alejandro Sánchez de Miguel[1] and Maruška Mole[2]

[1] **Dpto. Astrofísica y CC. de la Atmósfera, Facultad de CC. Físicas, Universidad Complutense de Madrid, Avda. Complutense s/n, E-28040 Madrid, Spain**
`fog@astrax.fis.ucm.es`

[2] **Center for Atmospheric Research, University of Nova Gorica, Vipavska 13, SI-5000 Nova Gorica, Slovenia**

Superfireballs are rare phenomena for which the reports are scarce and the estimation of their abundance has a huge margin of uncertainty. As a citizen science project we have gathered >500 reports from newspapers in the 1850-2000 period. This database shows how some superfireball abundances are constant during the period, though the reference newspapers have changed in the last two centuries. We have tentatively related some fireball sources to well-known meteor showers (Perseids, Geminids and Leonids), while superfireball sources may be related to minor or unknown showers, probably of asteroidal origin.

## 1 Introduction

Fireballs are a fascinating natural event that has been extensively reported and documented through the years. Chronicles from different cultures make account of the witnesses of these balls of fire, giving different explanations for these phenomena.

With the spreading of the alphabetization of the people and the apparition of the mass newspapers, the registration of meteor-related news grew dramatically. In the beginning of the 21$^{st}$ century the digitalization of these sources opens vast possibilities on the meteor research.

In particular, in the newspapers most of the events recorded are described 'as bright as full moon' or brighter. We call all of them 'superfireballs' though there is no consensus on the limit for such a category. Some authors set the limit at magnitude -17 while others give the name 'superbolide' to this group (Ceplecha et al., 1999). As a reference, the lower limit we consider is the brightness of full moon, so magnitude -13 corresponds to a 1 ton of TNT (4 GJ) event. For an average luminous efficiency and speed, that is a meteoroid of around 20 cm diameter or ~ 100 kg.

Superfireballs are too scarce for ground network statistics, and too dim for satellite detection in Earth orbit. These are the reasons for the gap that appears around the superfireballs category in the estimation of meteoroid influx at Earth (Brown et al., 2012).

Several authors have worked on fireball cataloguing, gathering reports for the last 20 centuries (Denning, 1912; Astapovic & Terentjeva, 1968; Terentjeva, 1989; Beech, 2006). However most of these archives are based on single-location and the number of superfireballs is limited. For instance, see Baggaley's research for the British Isles between 1900 and 1936 (Baggaley, 1977).

This research is based on the reports which appeared in newspapers in the last 150 years and were collected by volunteers in the frame of a citizen science project. The so-called 'Historical Fireballs' project has been made by a collaboration of PhD students, undergraduate students of the Faculty of Physics and members of the University astronomical club of the Complutense University ASAAF-UCM.

This project is intended as a beta-version of a future crowdsourcing project open to more countries and languages, but also is thought as an introduction to the science world for the students. The project has been presented in 2 national conferences (Sánchez de Miguel et al., 2014; Sánchez de Miguel et al., 2015) and this international conference. It has also appeared in the University outreach magazine (Zamora et al., 2015) and in a press released published by the University[1], which had wide mass-media coverage[2].

## 2 Method

As a citizen science project this project has involved many collaborators. In order to make the elaboration of databases easier Google Forms has been used. People can report the characteristics of the superfireballs found in the newspapers and other sources in this form. This form is the base for a future crowdsourcing of the work.

The research is based on several sources. Some of them are known journals as the New York Times and some Spanish journals for the XIX$^{th}$ century. To compare with the XX$^{th}$ century we used the Astrophysics Data System papers. For comparison purposes we collected reports from the British Association for the Advancement of Science (BAAS) and other catalogues. The main advantage of using newspapers is that we have a vast data sample for superfireballs avoiding some geographical bias (at least across longitudes), that data is already published and accessible and the news date back for more than 150 years.

---

[1] http://guaix.fis.ucm.es/historicalsuperbolides
[2] https://www.ucm.es/17-de-diciembre-de-2014

We have made a descriptive statistical analysis of the data, based on the mean and dispersion of the sample. We have assumed a normal distribution, considering a significant 2σ deviation from the rates along the year. We consider that non-parametrical analysis is essential for the next works, especially to avoid sampling interval bias. Our search of streams is based on meteor shower properties, with a bin size between 2 and 7 days, and the age of 'shower' (appearing for more than 50 years). We have tested this method on fireballs with good results.

The analysis is presented here by histograms of a number of events along the year (solar longitude 2015, in degrees). Also, we have considered the dates of the main meteor showers/storms (Lyrids, Perseids, Draconids, Orionids, Geminids and Quadrantids) and the progress of its event rate. We have developed software for this analysis of the data cataloged by volunteers of the citizen science project.

## 3  Databases

At the end of the compilation process there are a total of 2393 registries of fireballs obtained amongst five different databases.

We use as the main reference the New York Times journal. It has a digital record to find all journals. This fact enables a fast search using keywords, like meteorite, fireball, bolide, meteoritic stone, aerolite, meteorite fall, ball of fire, stonefall and similar terms in Spanish. At last, we have 377 fireballs between the years 1800 and 2000 from this sample (see *Figure 1*). We have as well 70 events from the Biblioteca Nacional (Spanish National Library) from 1850 to 1900. This database contains more than 100 Spanish journals and historical press.

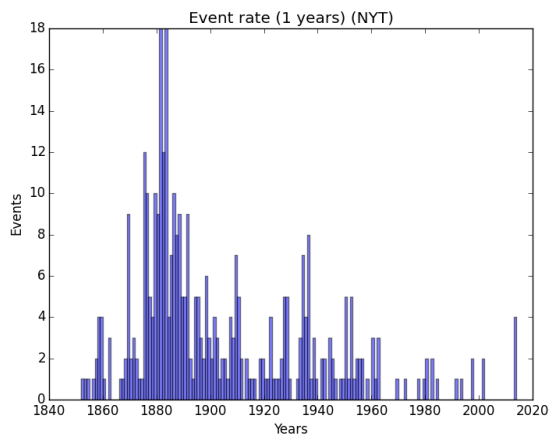

*Figure 1* – Histogram of the 1-year event rate distibution in the NYT database. It contains almost 400 superfireball reports until 2015, but most of them are part of the period 1850-1950.

In addition to newspaper stories we have entered some superfireballs from papers registered at the SAO/NASA Astrophysics Data System (ADS)[3]. From papers in this portal we have compiled 107 events in the range 1900-2000.

---

[3] http://adswww.harvard.edu/

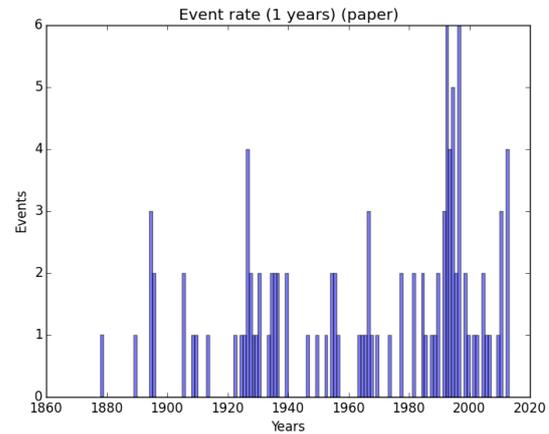

*Figure 2* – Histogram of the 1-year event rate distibution in the ADS database. More than 100 events are part of it. Most of them took place after 1950, complementing the NYT database.

For the purpose of comparing the method, software and statistical analysis, another database used is the British Association for the Advancement of Science (BAAS). We have 1634 entries of this sample from 400 A.D. to 2000. They are mainly fireballs so we expect to get similar results as other studies in order to validate ours.

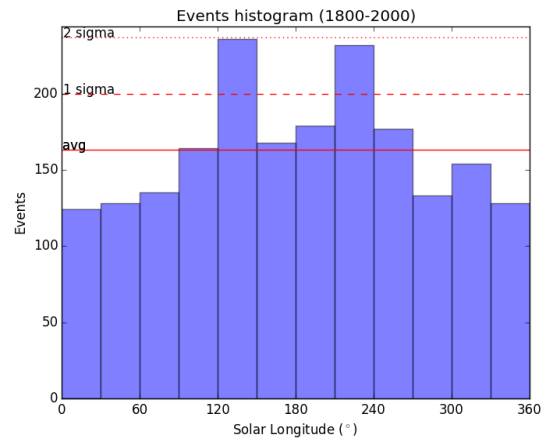

*Figure 3* – Histogram of data from all databases. It contains mainly fireballs from the BAAS catalogue, therefore we observe an annual trend, with a maximum around solar longitude 220º and the peaks of the Perseids and the Leonids as expected.

## 4  Results

### Superfireball rate along the decades and throughout the year

Through the decades the number of reported superfireballs is not constant, peaking around 1880 (see *Figure 4*). It is similar to the annual meteorite fall rate (Beech, 2002). It does not reflect a real trend, but a bias in the reports. This leads to the conclusion that our database of superfireballs is not adequate for studies of the absolute number of events per year.

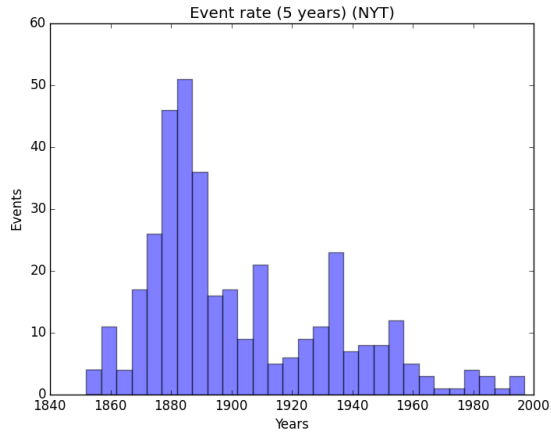

*Figure 4* – Histogram of the 5-year event rate distribution from 1840 on. The reports of superfireballs almost dissapear after the start of the Space Race.

However we can work with these relative values, as a representative sample and perform research on the number of events throughout the year. This number changes throughout the year. Most of the deviations from the mean are not large enough to be considered statistically significant. However some peaks appear while surveying the parameter space as explained above (bin size, age period) and may be superfireball streams (see *Figure 6*).

**Superfireballs and meteor showers**

In order to confirm the suitability of the method and tools developed, we use all available databases (dominated by fireballs, not superfireballs) and expect to find meteor showers peaks. The peaks corresponding to the Perseids and the Geminids are above the 2σ level, while other showers are over the 1σ level as well (*Figure 5*).

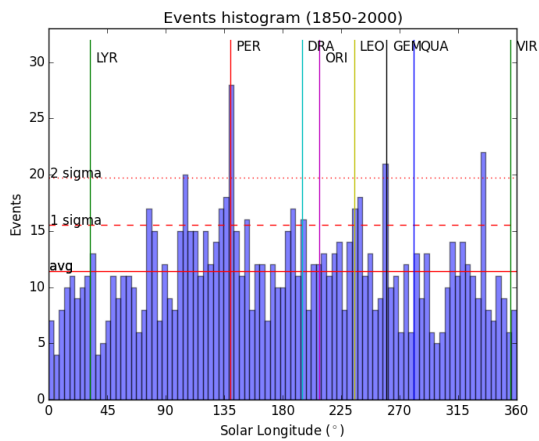

*Figure 5* – Histogram of all the databases. Fireball activity peaks match some of the major shower peak dates, represented as vertical lines.

However in case of just superfireballs from the NYT database meteor showers do not appear, which is in accordance with the lack of superfireballs observed during major meteor showers. But there are some other peaks, which could be superfireball streams (see *Figure 6*).

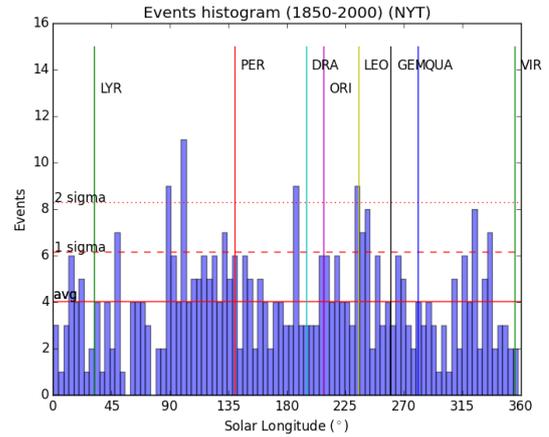

*Figure 6* – Histogram of the NYT database for the 1850–2000 period with the peak date of major showers represented. They do not match the dates with stronger superfireball activity.

**Possible superfireball streams**

The existence of the superfireball streams requires a model to explain their nature and possible characteristics. For the moment our search for them is restricted to a certain range of length (bin size between 2 and 7 days) and the age of the 'shower' (more than 50 years, up to almost 200 years). However some other real streams may be blurred due to the selection of these parameters.

The histograms in *Figures 7–9* show one possible representation of event distribution for the selected bin size. We have identified these streams with the center date (epoch 2015) of the bins with larger significance.

For the period 1850–2000 we have found three streams with maximum strength in 2-day bins:

- July 3 ($\lambda_\odot = 101°$)
- October 2 ($\lambda_\odot = 188°$)
- February 22 ($\lambda_\odot = 333°$)

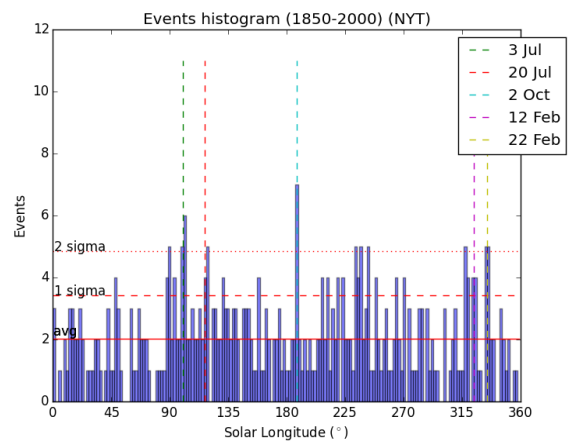

*Figure 7* – Histogram of the NYT database for the 1850–2000 period with 2-day bins. Three of the streams found in the survey are clearly present, and the other two have some significancy.

However others appear only for longer bin sizes, for 7-day bin in the XIX[th] century with center in:

- July 20 ($\lambda_\odot = 117°$)

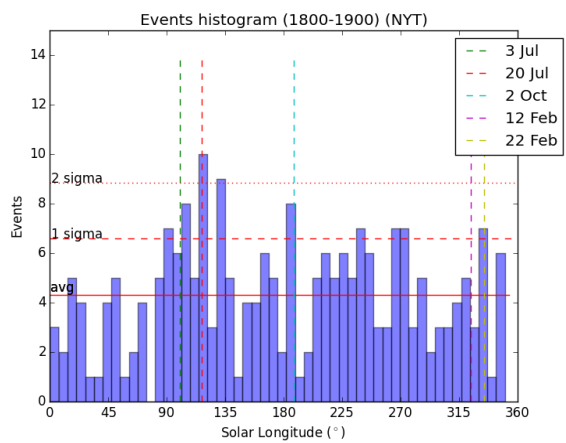

*Figure 8* – Histogram of the NYT database for the 1800–1900 period with 7-day bins. The 20[th] July stream is visible with a peak not far from that date.

Another stream is only observed in the XX[th] century, using a 3-day bin centered in:

- February 12 ($\lambda_\odot$=323°)

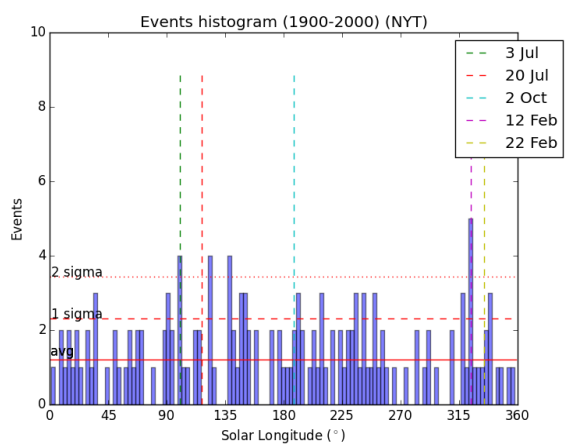

*Figure 9* – Histogram of the NYT database for the 1900–2000 period with 3-day bins. The 12[th] February stream stands out above the average.

Therefore we think that these superfireball streams are real. They could be short-lived (50–100 years) or even shorter, though others appear through both centuries. Some of these streams are narrow (2-day bins) while others may be quite wide (7-day bins).

## 5  Future work

As a citizen science project the next step is to open it to more collaborators using crowdsourcing environments like other projects of GUAIX group[4]. It will help to complete the databases and increase the number of sources (more countries, more languages).

Regarding the analysis we plan to work further in the statistics and start to pinpoint the radiants in order to characterize the proposed superfireball streams. To find the possible radiant we plan to combine geographical data and time of observation data for all the fireballs.

The characterization of them would help to determine their origin scenario. Comets do not release enough of such big boulders (Baranov and Smirnov, 2005), so asteroidal collision is the most likely mechanism for the formation of these streams (Fujiwara et al., 1989). However the lack of smaller particles requires a complex delivery mechanism, with mass sorting. The duration of these 'showers' would help to constrain the age of the stream.

## 6  Conclusion

The superfireball reports gathered for the period covering the last 150 years show statistically significant overabundance peaks for some periods of the year. During these periods we find no evidence of fireballs/meteor showers.

These intervals are centered in (dates in epoch 2015):

- July 3 ($\lambda_\odot$=101°)
- July 20 ($\lambda_\odot$=117°)
- October 2 ($\lambda_\odot$=188°)
- February 12 ($\lambda_\odot$=323°)
- February 22 ($\lambda_\odot$=333°)

## Acknowledgments

The authors thank all the volunteers that have contributed to the elaboration of the databases used for this research, especially the members of the university astronomical club ASAAF-UCM[5]. F.O. and A.S. thank the Spanish Meteor Network (SPMN) researchers for their support and all the fruitful discussions.

---

[4] https://guaix.fis.ucm.es/ISS_cities_at_night

[5] http://www.asaaf.org/